\documentclass{article}

\usepackage{aimc2025}

\usepackage[utf8]{inputenc} 
\usepackage[T1]{fontenc}    
\usepackage{hyperref}       
\hypersetup{
  colorlinks   = true, 
  urlcolor     = blue, 
  linkcolor    = black, 
  citecolor   = black 
}
\usepackage{url}            
\usepackage{booktabs}       
\usepackage{amsfonts}       
\usepackage{nicefrac}       
\usepackage{microtype}      
\usepackage{graphicx}       

\usepackage{amsmath}

\usepackage{xcolor}
\usepackage{siunitx}
\usepackage{float}
\usepackage{tikz}
\usetikzlibrary{shapes.geometric, arrows, positioning}
\usepackage{pgfplots}
\usepgfplotslibrary{groupplots,dateplot}

\definecolor{highlight1}{HTML}{0077BB} 
\definecolor{highlight2}{HTML}{EE7733} 
\definecolor{highlight3}{HTML}{009988} 
\definecolor{highlight4}{HTML}{CC3311} 
\definecolor{highlight5}{HTML}{AA4499} 
\definecolor{highlight6}{HTML}{EECC66} 

\def\note[#1#2#3]{#1\if b#2$\flat_#3$\else\if#2##$\sharp_#3$\else$_#2$\fi\fi}

\title{Exploring Procedural Data Generation for Automatic Acoustic Guitar Fingerpicking Transcription}

\author{%
  Sebastian Murgul \\
  Klangio GmbH \\
  Karlsruhe, Germany \\
  \texttt{sebastian.murgul@klang.io} \\
  \And
  Michael Heizmann \\
  Karlsruhe Institute of Technology \\
  Karlsruhe, Germany \\
  \texttt{michael.heizmann@kit.edu} \\
}

\begin{document}

\maketitle

\begin{abstract}
  Automatic transcription of acoustic guitar fingerpicking performances remains a challenging task due to the scarcity of labeled training data and legal constraints connected with musical recordings. This work investigates a procedural data generation pipeline as an alternative to real audio recordings for training transcription models. Our approach synthesizes training data through four stages: knowledge-based fingerpicking tablature composition, MIDI performance rendering, physical modeling using an extended Karplus-Strong algorithm, and audio augmentation including reverb and distortion. We train and evaluate a CRNN-based note-tracking model on both real and synthetic datasets, demonstrating that procedural data can be used to achieve reasonable note-tracking results. Finetuning with a small amount of real data further enhances transcription accuracy, improving over models trained exclusively on real recordings. These results highlight the potential of procedurally generated audio for data-scarce music information retrieval tasks.
\end{abstract}

\section{Introduction}\label{sec:introduction}
Automatic guitar transcription is a longstanding challenge in the field of Music Information Retrieval (MIR), particularly for polyphonic and expressive styles such as fingerpicking. Fingerpicking is a widely used guitar technique in which strings are plucked individually using the fingers or a plectrum, resulting in complex rhythmic and harmonic textures. Transcribing such performances into symbolic representations like tablature or MIDI remains difficult due to both musical and practical constraints.

While substantial progress has been made in piano transcription using deep learning methods, automatic guitar transcription has received comparatively less attention. One major limitation is the scarcity of high-quality annotated data. \citet{xi2018guitarset} introduced \textit{GuitarSet}, one of the few datasets with detailed note-level annotations for real guitar recordings. It includes 360 performances across various genres, tempi, and styles, recorded using a hexaphonic pickup system. However, the dataset is relatively small and lacks annotations for expressive playing techniques such as slides, bends, or hammer-ons.
Wiggins et al.~used GuitarSet to train a Convolutional Neural Network (CNN) to transcribe the audio into string-wise MIDI representations \citep{wiggins2019guitar}.
To address symbolic data scarcity, \citet{sarmento2021dadagp} proposed \textit{DadaGP}, a large-scale corpus of over 26,000 guitar pieces in GuitarPro format, accompanied by a tokenizer for sequence modeling. Although extensive, DadaGP consists solely of symbolic data and contains no audio. Building on this, \citet{zang2024synthtab} introduced \textit{SynthTab}, which renders audio from GuitarPro files using commercial VST instruments, enabling training of transcription models on paired symbolic and audio data. However, SynthTab depends on proprietary tools and does not provide expressive performance annotations.
Recent work has also investigated learning transcription models without relying on detailed annotations. \citet{wiggins2020towards} proposed a weakly supervised framework for acoustic guitar transcription that leverages unpaired tablature and audio data. Their system attempts to align symbolic representations with real guitar recordings using heuristic constraints, bypassing the need for frame-level note labels. While promising, such methods still face limitations in accurately modeling articulation and timing nuances.
Moreover, several studies have explored guitar synthesis as a tool for data generation and interaction. The Karplus-Strong algorithm \citep{karplus1983digital} and its extensions \citep{jaffe1983extensions, sullivan1990extending} remain popular due to their efficiency and ability to simulate the dynamics of plucked strings. Convolution-based body modeling methods, such as those by \citet{lopez2008acoustic}, enhance realism by simulating guitar body resonance through impulse responses.
A more complex set of numerical simulation tools has been introduced by \citet{tahvanainen2019numerical} to leverage the virtual analysis of guitar mode frequencies, frequency responses, and radiation efficiency in an industrial context.
More recently, \citet{jonason2023ddsp} introduced a DDSP-based neural synthesis approach for generating polyphonic guitar audio from string-wise MIDI input.
In 2024, Bilbao et al.~presented a new, efficient, and numerically stable method for real-time guitar synthesis that models complex string dynamics and nonlinear interactions without requiring computationally expensive iterative solvers \citep{bilbao2024real}.

Despite these advances, most current transcription systems are trained using supervised methods, requiring large datasets of audio paired with precise note annotations. Legal restrictions connected with the use of copyrighted music recordings or MIDI files can make it difficult or expensive to acquire such datasets, particularly for commercial applications.

Furthermore, by using a procedural data generation process, it becomes possible to enable the transcription of more creative playing styles and techniques. This is achieved by creating a more balanced data distribution through the targeted generation of underrepresented musical features. As a result, AI transcription models trained on such data may better capture and notate creative or unconventional performances that are typically underrepresented in real-world datasets, thus ensuring that these artistic expressions are preserved and not lost.

In this work, we propose a fully procedural data generation pipeline as an alternative to using real recordings for training. Our method combines knowledge-based composition of fingerpicking tablature, expressive MIDI performance rendering, physical modeling using an extended Karplus-Strong algorithm, and audio augmentation simulating room acoustics and recording imperfections. This approach enables the scalable creation of musically coherent, expressive, and annotated training data, offering a promising solution for data-scarce transcription scenarios.
\section{Procedural data generation}\label{sec:methodology}

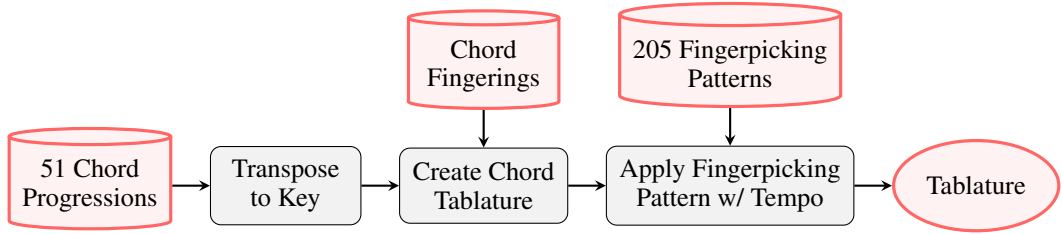
\begin{figure}[tb]
  \begin{tikzpicture}[
    auto,
    node distance=0.5cm,
    roundnode/.append style={ellipse, draw=red!60, fill=red!5, very thick, minimum width = 2cm, minimum height = 1.2cm, align=center},
    process/.append style={rectangle, rounded corners, minimum width=2cm, minimum height=1cm, align=center, draw=black, fill=black!5, inner sep=5pt},
    database/.append style={cylinder, minimum width=2cm, minimum height=1cm, align=center, draw=red!60, fill=red!5, very thick, inner sep=5pt, shape border rotate=90, aspect=0.1},
    arrow/.style={thick,->,>=stealth}]

  \node (progression)[database] {51 Chord \\Progressions};
  \node (transpose) [process, right = of progression] {Transpose \\to Key};
  \node (tablature) [process, right = of transpose] {Create Chord\\ Tablature};
  \node (fingerings) [database, above = of tablature] {Chord \\Fingerings};
  \node (apply-pattern) [process, right = of tablature] {Apply Fingerpicking \\Pattern w/ Tempo};
  \node (patterns) [database, above = of apply-pattern] {205 Fingerpicking\\Patterns};
  \node (output) [roundnode, right = of apply-pattern] {Tablature};

  \draw [arrow] (progression) -- (transpose);
  \draw [arrow] (transpose) -- (tablature);
  \draw [arrow] (fingerings) -- (tablature);
  \draw [arrow] (tablature) -- (apply-pattern);
  \draw [arrow] (patterns) -- (apply-pattern);
  \draw [arrow] (apply-pattern) -- (output);

\end{tikzpicture}
  \centering
  \caption{Flow chart of the fingerpicking tablature sampling process.}
  \label{fig:fingerpicking_sampling}
\end{figure}

To address the scarcity of labeled training data for acoustic guitar transcription, particularly under legal constraints associated with copyrighted recordings, we propose a procedural data generation pipeline. The goal of this system is to synthesize large-scale, musically coherent, and expressive training data entirely through algorithmic means, eliminating the dependency on real audio or annotated MIDI files.

The key requirements for such a pipeline include: musical validity, ensuring that generated content adheres to realistic harmonic and rhythmic structures; expressive realism, capturing the imperfections and tonal nuance of human performances; computational efficiency, allowing large datasets to be created without introducing training bottlenecks.

Our pipeline consists of four main stages. It begins with the composition of fingerpicking tablatures, where chord progressions are combined with stylistic picking patterns to create musically plausible pieces. These tablatures are then transformed into expressive MIDI sequences using a performance rendering step that introduces timing and pitch variability to emulate human playing. The MIDI is rendered into audio using an extended Karplus-Strong algorithm, which simulates the physical behavior of plucked guitar strings. Finally, an audio augmentation stage applies effects such as distortion, filtering, reverb, and noise to emulate diverse recording environments and increase the acoustic diversity of the dataset.

This modular design enables the generation of rich, annotated audio data suitable for training transcription models in data-scarce settings, and provides a scalable alternative to curated real-world datasets.

\subsection{Fingerpicking tablature sampling}

The data generation process begins with the creation of synthetic fingerpicking guitar tablatures, as illustrated in Figure~\ref{fig:fingerpicking_sampling}. This step aims to produce musically realistic and stylistically diverse compositions suitable for rendering into audio and training transcription models.

To achieve this, we use a curated database containing $51$ chord progressions written in functional harmony and $205$ fingerpicking patterns arranged on a 16th-note rhythmic grid. The picking patterns are inspired by classic fingerstyle repertoire as cataloged by \citet{manzi2000fingerpicking}, and are encoded using the \emph{PIMA} system; an abbreviation derived from Spanish finger names: \emph{P} (pulgar, thumb), \emph{I} (índice, index), \emph{M} (medio, middle), and \emph{A} (anular, ring). This notation allows us to specify which right-hand finger plucks which string in a given rhythmic slot.

Each generated tablature is created through a multi-step sampling process. First, a chord progression is randomly selected and transposed to a randomly chosen key. Each chord is then mapped to a specific fingering using a pre-defined lookup table. Next, a fingerpicking pattern is sampled from the database and applied over the chord progression at a randomly selected tempo. Patterns are available in various time signatures (4/4, 3/4, 6/8, and 12/8), and can be applied to any chord containing at least four active strings.

The plucking instructions rely on string position encoding: positive indices count downward from the highest-pitched string (high \note[E4]), while negative indices count upward from the lowest-pitched string (low \note[E2]). This system ensures compatibility with complex fingerings and variable chord voicings.
Theoretically, it can be applied to any guitar tuning, but here we focus on the standard tuning.

\begin{figure}[htb]
  \centering
  \includegraphics[width=0.99\textwidth]{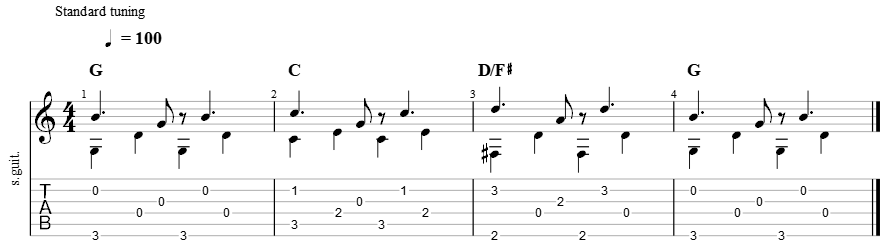}
  \caption{Example tablature generated by the proposed fingerpicking generator using a Travis picking pattern from the pattern database.}
  \label{fig:example_composition}
\end{figure}

This knowledge-based approach enables the generation of an extensive number of musically coherent and stylistically diverse fingerpicking pieces, without the need for manually annotated datasets. An example output, using a Travis picking pattern, is shown in Figure~\ref{fig:example_composition}.

Tablatures are handled programmatically using the \texttt{PyGuitarPro} library~\citep{abakumov2023pyguitarpro}, which supports reading, editing, and exporting Guitar Pro files. This allows for direct visualization and editing of generated tablatures using standard notation software.

\subsection{MIDI performance creation}

Once the fingerpicking tablatures are generated, they are converted into expressive MIDI performances that emulate human playing. This step bridges the gap between symbolic composition and audio synthesis by introducing temporal and pitch-level variations through a process known as humanization.

Each tablature is translated into a sequence of MIDI-style note events, where timing and pitch characteristics are perturbed to increase realism and variability. Timing deviations are introduced by adding random jitter to both the note onset and offset, with a maximum deviation of $\SI{10}{\percent}$ of the note's nominal duration. This simulates subtle imperfections in timing typically present in human performances.

To enhance pitch diversity, we apply probabilistic pitch perturbations that introduce melodic variations and bass runs into the underlying fingerpicking tablature. For each note, there is an $\SI{80}{\percent}$ chance the original pitch is retained. With a $\SI{5}{\percent}$ chance each, the pitch is shifted up or down by one or two semitones. Although this reduces the strict fidelity of the original composition, it introduces melodic contour, particularly on higher strings, that has been found to improve transcription performance in solo guitar passages.

This performance-level variation enhances the acoustic diversity of the training data, helping the transcription model generalize better to expressive or imperfect recordings.

\subsection{Audio rendering}

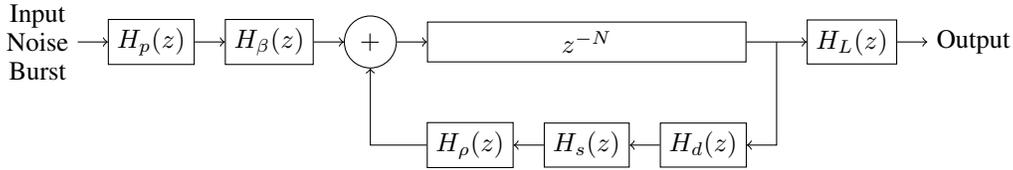
\begin{figure}[ht]
  \begin{tikzpicture}
  \node [align=center] (input) {Input\\Noise\\Burst};
  \node [draw, rectangle, right=0.4cm of input] (Hp) {$H_p(z)$};
  \node [draw, rectangle, right=0.4cm of Hp] (Hb) {$H_\beta(z)$};
  \node [circle, draw, right=0.4cm of Hb] (sum) {$+$};
  \node [draw, rectangle, right=0.4cm of sum, text width=4cm, align=center] (delay) {$z^{-N}$};
  \node [draw, rectangle, right=0.8cm of delay] (HL) {$H_L(z)$};
  \node [right=0.4cm of HL] (output) {Output};

  \node [draw, rectangle, below=0.8cm of delay] (Hs) {$H_s(z)$};
  \node [draw, rectangle, left=0.4cm of Hs] (Hr) {$H_\rho(z)$};
  \node [draw, rectangle, right=0.4cm of Hs] (Hd) {$H_d(z)$};

  \path (delay) -- (HL) coordinate [midway] (cdelay);

  \draw [->] (input) -- (Hp);
  \draw [->] (Hp) -- (Hb);
  \draw [->] (Hb) -- (sum);
  \draw [->] (sum) -- (delay);
  \draw [->] (delay) -- (HL);
  \draw [->] (HL) -- (output);

  \draw [->] (cdelay) |- (Hd);
  \draw [->] (Hd) -- (Hs);
  \draw [->] (Hs) -- (Hr);
  \draw [->] (Hr) -| (sum);
\end{tikzpicture}
  \centering
  \caption{Flow chart of the extended Karplus-Strong synthesis method, adapted from \citet{jaffe1983extensions}.}
  \label{fig:karplus_strong}
\end{figure}

To synthesize audio from MIDI sequences, we employ an extended Karplus-Strong algorithm~\citep{jaffe1983extensions}, which models the behavior of a vibrating string using a delay line and a series of digital filters. This physically inspired method is both computationally efficient and highly controllable, making it well suited for scalable procedural data generation. An overview of the synthesis process is illustrated in Figure~\ref{fig:karplus_strong}.

The synthesis begins with an excitation signal, typically a burst of filtered white noise, which is fed into a recursive delay loop. The delay length $N$ corresponds to the pitch period in samples (twice the simulated string length). The loop includes several digital filters that simulate various physical properties of the string:

\begin{align*}
  H_p(z)     & = \frac{1 - p}{1 - p z^{-1}}                  &  & \text{(pick-direction lowpass filter)}                                             \\
  H_\beta(z) & = 1 - z^{-|\beta N + 1/2|}                    &  & \text{(pick-position comb filter), } \beta \in (0,1)                               \\
  H_d(z)     & = \text{string-damping filter}                &  & \text{(must satisfy } |H_d\left(e^{j\omega T}\right)| \leq 1\text{ for stability)} \\
  H_s(z)     & = \text{string-stiffness allpass filter}      &  & \text{(simulating dispersion)}                                                     \\
  H_\rho(z)  & = \frac{\eta(N) - z^{-1}}{1 - \eta(N) z^{-1}} &  & \text{(tuning allpass filter)}                                                     \\
  H_L(z)     & = \frac{1 - R_L}{1 - R_L z^{-1}}              &  & \text{(dynamic-level lowpass filter)}
\end{align*}

Each filter plays a distinct role in modeling the acoustic behavior of the string: $H_p(z)$ adjusts spectral roll-off based on pick direction. The parameter $p \in [0, 1]$ defines the pole position in the lowpass filter, with $p = 0 $ for one direction and $p \in (0, 1]$, for the opposite. $H_\beta(z)$ simulates the effect of pick position along the string, controlled by the normalized position parameter $\beta$. $H_d(z)$ applies damping to simulate energy decay over time. For stability, the frequency response must satisfy $|H_d(e^{j\omega T})| \leq 1$. $H_s(z)$ models stiffness-related dispersion using an allpass filter with multiple poles and zeros. $H_\rho(z)$ allows fine pitch adjustment via a fractional-delay allpass filter. The coefficient $\eta \in [-1/11, 2/3]$ adjusts the effective delay in the range $[0.2, 1.2]$ samples. $H_L(z)$ simulates dynamic-level dependent brightness, with
$R_L = e^{-\pi L T}$, where $L$ is the desired bandwidth in Hz and $T$ is the sampling interval.

To enhance diversity and realism, we randomly sample the following synthesis parameters for each note:

\begin{align*}
  \text{Amplitude}~A      & \in [0.2, 1.3]                     \\
  \text{Brightness}~\beta & \in [0.1, 0.9]                     \\
  \text{Level}~L          & \in [0.1, 0.9]                     \\
  \text{Pick Position}~p  & \in [0.1, 0.9]                     \\
  \text{Detune}~\delta    & \in [-0.49, 0.49]~\text{semitones}
\end{align*}

Detuning is applied by offsetting the MIDI pitch $m$ before converting to frequency. The fundamental frequency $f_0$ is calculated as

\begin{equation}
  f_0 = 440 \cdot 2^{\frac{m + \delta - 69}{12}},
\end{equation}

where $\delta$ is the detune value in fractional semitones. The delay length is then computed as

\begin{equation}
  N = \frac{f_s}{f_0},
\end{equation}

with $f_s$ being the sampling rate of $\SI{16}{\kilo\hertz}$. This method introduces subtle, realistic pitch variations between notes, mimicking tuning inconsistencies found in real guitar performances and improving model robustness.

By combining physical modeling with randomized parameter modulation, the extended Karplus-Strong synthesis engine produces highly expressive, diverse, and controllable audio signals.

\subsection{Audio augmentation}

To enhance realism and bridge the gap between synthetic and real-world recordings, we apply a post-processing audio augmentation pipeline using the \texttt{Pedalboard} library by \citet{sobot2023pedalboard}. This stage introduces acoustic variability of the recording equipment and environment, aiming to improve the robustness of the transcription model when deployed on non-synthetic data.

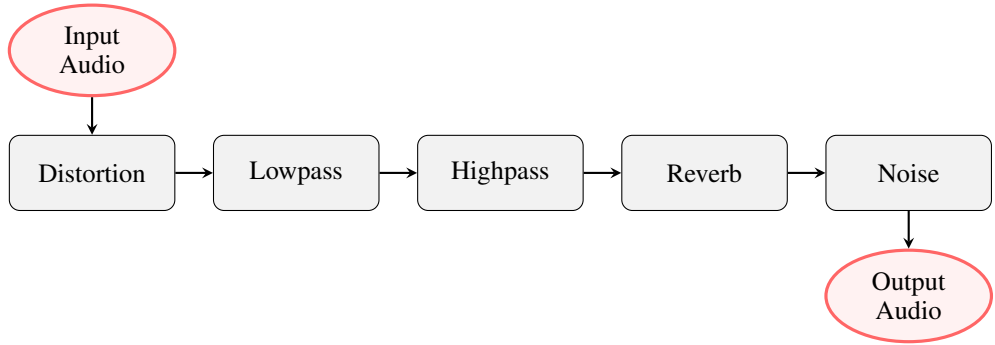
\begin{figure}[ht]
  \begin{tikzpicture}[
    auto,
    node distance=0.5cm,
    roundnode/.append style={ellipse, draw=red!60, fill=red!5, very thick, minimum width = 2.2cm, minimum height = 1.2cm, align=center},
    process/.append style={rectangle, rounded corners, minimum width=2.2cm, minimum height=1cm, align=center, draw=black, fill=black!5, inner sep=5pt},
    arrow/.style={thick,->,>=stealth}]

  \node (input-audio)[roundnode] {Input\\ Audio};
  \node (distortion) [process, below = of input-audio] {Distortion};
  \node (lp) [process, right = of distortion] {Lowpass};
  \node (hp) [process, right = of lp] {Highpass};
  \node (reverb) [process, right = of hp] {Reverb};
  \node (noise) [process, right = of reverb] {Noise};
  \node (output-audio) [roundnode, below = of noise] {Output\\ Audio};

  \draw [arrow] (input-audio) -- (distortion);
  \draw [arrow] (distortion) -- (lp);
  \draw [arrow] (lp) -- (hp);
  \draw [arrow] (hp) -- (reverb);
  \draw [arrow] (reverb) -- (noise);
  \draw [arrow] (noise) -- (output-audio);

\end{tikzpicture}
  \centering
  \caption{Flow chart of the audio augmentation pipeline applied to synthesized fingerpicking recordings.}
  \label{fig:audio_augmentation}
\end{figure}

As illustrated in Figure~\ref{fig:audio_augmentation}, the augmentation chain includes distortion, highpass and lowpass filtering, convolutional reverb, and additive noise. Each effect is applied independently with a $\SI{50}{\percent}$ probability, and its parameters are randomly modulated to ensure diverse and plausible output variations. Distortion is applied with a randomly sampled drive level in the range of $1$ to $\SI{4}{\decibel}$, simulating nonlinear saturation such as pickup overdrive or analog warmth. Lowpass filters are configured with cutoff frequencies uniformly sampled between $\SI{1.5}{\kilo\hertz}$ and $\SI{8}{\kilo\hertz}$, while highpass filters use cutoff frequencies in the range of $\SI{50}{\hertz}$ to $\SI{500}{\hertz}$, mimicking tonal shaping introduced by different microphones or hardware. Reverb is added using convolution with impulse responses, where the virtual room size is randomly chosen between $0.25$ and $1.0$, simulating a range from small practice spaces to large halls. Finally, white noise is injected at a randomly selected signal-to-noise ratio (SNR) between $\SI{30}{\decibel}$ and $\SI{50}{\decibel}$, representing environmental or equipment noise often present in non-studio conditions.

This controlled randomness introduces meaningful variability into the dataset without significantly altering the underlying musical content. As a result, the augmented audio more accurately reflects the diversity and imperfections of real acoustic guitar recordings, leading to better generalization in downstream transcription tasks.
Although designed for synthesized audio, the augmentation pipeline is generalizable and can also be applied to VST-generated data or real recordings.

\section{Experimentation setup}
\label{sec:experiments}

This section details the model architecture, dataset, and training configuration used in our experiments.

\subsection{Model}

We adopt the CRNN-based \textit{Onsets and Frames} (OaF) model proposed by \citet{hawthorne2018onsets} as the core architecture for note tracking. Despite the emergence of more recent approaches such as transformer-based models~\citep{gardner2022mt3} and regression-based networks~\citep{riley2024high}, we selected OaF for its training efficiency, simplicity, and strong baseline performance.

The model is adapted for acoustic guitar transcription by modifying its output dimensionality and tuning hyperparameters accordingly. Input audio is resampled to $\SI{16}{\kilo\hertz}$ and converted into a log-scaled Mel spectrogram using a window size of $2048$ samples and a hop size of $512$. This results in a time-frequency representation with $229$ frequency bins, starting from a minimum frequency of $\SI{30}{\Hz}$.

\begin{figure}[ht]
\tikzstyle{block} = [rectangle, draw, text width=10em, text centered, minimum height=1.6em]

\begin{tikzpicture}
  \coordinate (dummy);
  \node [block, above=1cm of dummy] (mel) {Log Mel-Spectrogram \\($B \times N \times 229$)};

  \node [block, right=1cm of dummy] (convl) {Conv Stack};
  \node [block, below=.5cm of convl] (fcl1) {FC Sigmoid};
  \node [block, below=.5cm of fcl1] (bilstml1) {BiLSTM};
  \node [block, below=.5cm of bilstml1] (fcl2) {FC Sigmoid};
  \node [block, below=.5cm of fcl2] (framepred) {Frame Predictions ($B \times N \times 49$)};
  \node [block, below=.5cm of framepred] (frameloss) {Frame Loss};

  \node [block, left=1cm of dummy] (convr) {Conv Stack};
  \node [block, below=.5cm of convr] (bilstmr) {BiLSTM};
  \node [block, below=.5cm of bilstmr] (fcr1) {FC Sigmoid};
  \node [block, below=.5cm of fcr1] (onsetpred) {Onset Predictions ($B \times N \times 49$)};
  \node [block, below=.5cm of onsetpred] (onsetloss) {Onset Loss};

  \path (onsetpred) -- (onsetloss) coordinate [midway] (conset);
  \path (bilstml1) -- (fcr1) coordinate [midway] (cbilstm);
  \path (bilstml1) -- (fcl1) coordinate [midway] (cfcl);

  \draw [->] (mel) -- (convl);
  \draw [->] (convl) -- (fcl1);
  \draw [->] (fcl1) -- (bilstml1);
  \draw [->] (bilstml1) -- (fcl2);
  \draw [->] (fcl2) -- (framepred);
  \draw [->] (framepred) -- (frameloss);

  \draw [->] (mel) -- (convr);
  \draw [->] (convr) -- (bilstmr);
  \draw [->] (bilstmr) -- (fcr1);
  \draw [->] (fcr1) -- (onsetpred);
  \draw [->] (onsetpred) -- (onsetloss);

  \draw [->] (conset) -| (cbilstm) |- ([xshift=-0.2cm]cfcl) -| ([xshift=-0.2cm]bilstml1.north);
\end{tikzpicture}
  \centering
  \caption{Flow chart of the Onsets and Frames model architecture by \citet{hawthorne2018onsets} adapted for guitar note-tracking.}
  \label{fig:model_architecture}
\end{figure}
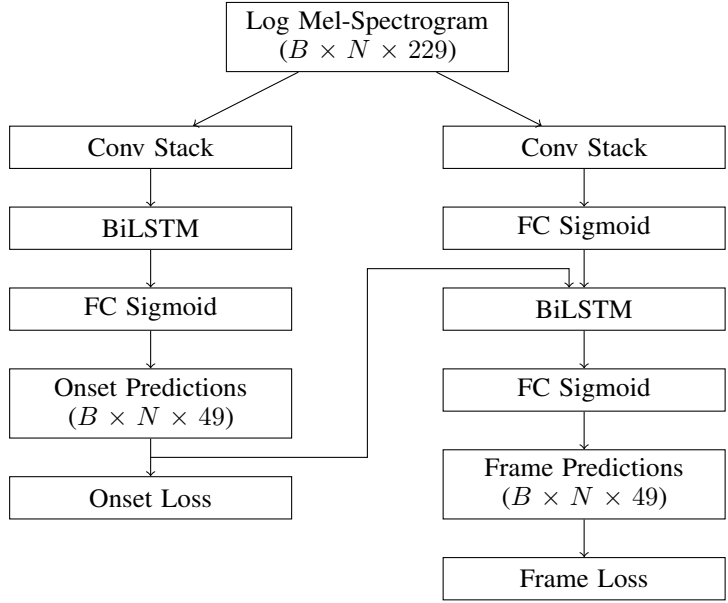

The architecture is visualized in Figure \ref{fig:model_architecture} and consists of two parallel processing branches, one for note onsets and one for sustained frames, each producing a $(B \times N \times 49)$ piano roll representation. Here, $B$ denotes the batch size and $N$ the dynamically set number of time frames. The pitch range spans from \note[E2] (MIDI $40$) to \note[E6] (MIDI $88$), covering standard acoustic guitar tuning.

Each branch begins with a convolutional stack of three $3 \times 3$ layers, each followed by batch normalization and ReLU activation. Max pooling and dropout are applied after the second and third convolutional layers. The resulting feature maps are passed through a fully connected layer that compresses the embedding to $256$ dimensions. This embedding is then fed into a bidirectional LSTM (BiLSTM), followed by a final fully connected layer with sigmoid activation to output note probabilities.

For training, we use binary cross-entropy losses for both the onset and frame outputs. The total loss is computed as the sum of these two components.

\subsection{Datasets}

We use the GuitarSet dataset~\citep{xi2018guitarset} to train our baseline checkpoints and evaluate our models. GuitarSet consists of $360$ annotated recordings, including both solo and accompaniment (comping) performances from six guitarists. Recordings are captured using a hexaphonic pickup, enabling semi-automatic note-level annotations.

To ensure subject independence during evaluation, we use recordings from one guitarist (the first subject) as the test set, and the remaining five as the training set. Evaluation metrics are reported separately for solo and accompaniment subsets within the test data.

\subsection{Training}

All models are trained using the Adam optimizer with an initial learning rate of $6 \times 10^{-4}$. Gradient clipping with a threshold of $3$ is applied to stabilize training. We train for 10,000 steps using a batch size of $8$. On an NVIDIA Tesla V100 GPU, each full training run takes approximately $\SI{2}{\hour}$.

For finetuning experiments, we reduce the learning rate to $6 \times 10^{-5}$ and halve the batch size to $4$, allowing for more fine-grained updates on the real data.
\section{Results}
\label{sec:results}

To assess the quality of the MIDI transcriptions, we report note-level precision, recall, and F1-Score, using a $\SI{50}{ms}$ tolerance window in accordance with the \texttt{mir\_eval} library~\citep{raffel2014mireval}.

\subsection{Baseline evaluations}

\begin{table}
  \caption{Baseline comparison of the note Precision (P), Recall (R), and F1-Score (F1) results in percent for different audio sources on the GuitarSet test split. We apply metrics to the full test split, as well as to the accompaniments and solos individually.}
  \label{tab:baseline_results}
  \centering
  \begin{tabular}{l|ccc|ccc|ccc}
    \toprule
    \textbf{Audio Source} & \multicolumn{3}{c|}{\textbf{Full}} & \multicolumn{3}{c|}{\textbf{Comp}} & \multicolumn{3}{c}{\textbf{Solo}}                                                                                 \\
    \cmidrule{2-10}
                          & \textbf{P}                         & \textbf{R}                         & \textbf{F1}                       & \textbf{P} & \textbf{R} & \textbf{F1} & \textbf{P} & \textbf{R} & \textbf{F1} \\
    \midrule
    Real Recordings       & $92.05$                            & $73.84$                            & $81.42$                           & $90.75$    & $63.02$    & $74.15$     & $93.35$    & $84.66$    & $88.70$     \\
    VST Synthesis         & $84.87$                            & $63.88$                            & $71.58$                           & $87.36$    & $51.23$    & $64.17$     & $82.37$    & $76.52$    & $78.98$     \\
    Karplus-Strong        & $80.62$                            & $60.83$                            & $68.22$                           & $79.02$    & $48.56$    & $59.41$     & $82.23$    & $73.09$    & $77.02$     \\
    \bottomrule
  \end{tabular}
\end{table}

We begin by evaluating transcription performance on three different training sources: real audio recordings, VST-synthesized audio, and audio synthesized using the extended Karplus-Strong model. Each model is trained for $10,000$ steps to ensure a fair and consistent comparison across sources. Table~\ref{tab:baseline_results} summarizes the results.

VST Audio rendering is performed using the \texttt{DAWDreamer} Python library~\citep{braun2021dawdreamer} and Ample Sound's sample-based virtual guitar instruments\footnote{\url{https://amplesound.net/en/index.asp}}, following a methodology similar to SynthTab~\citep{zang2024synthtab}. Our Karplus-Strong synthesis setup includes all parameter modulations and audio augmentations described in Section~\ref{sec:methodology}.

As expected, training on real recordings yields the highest scores for both accompaniment and solo tracks. Across all three audio sources, solos consistently outperform accompaniment recordings, likely due to the higher note density in comping tracks. VST training data results in an F1-Score approximately $\SI{12}{\percent}$ lower than real recordings. The gap can be attributed to the VST's lack of expressive imperfections and recording noise present in amateur performances.

Despite being fully synthetic and not derived from any real guitar audio, the Karplus-Strong synthesis achieves performance comparable to the VST baseline. In contrast, VST rendering such as with Ample Sound requires running within a headless DAW environment like DAWDreamer braun2021dawdreamer, which not only introduces overhead and slows down the generation process but also complicates the implementation of the procedural data generation pipeline. Furthermore, most plugins are only available for Windows and macOS, making them unsuitable for Linux-based workflows. Our method runs natively and efficiently on Linux systems with minimal resource demands, making it well suited for scalable procedural data generation.

\subsection{Effect of synthesis modulation} \label{sec:results_modulation}

To assess the impact of parameter modulation in the Karplus-Strong synthesis, we trained models on audio generated from the JAMS annotation files of the GuitarSet training split using both static and modulated parameters (without further audio augmentation). Default values were: amplitude = 1, brightness = 0.5, level = 0.2, position = 0.5, and no detuning.

As shown in Table~\ref{tab:synthesis_results}, the static version yields high precision but poor recall, indicating overfitting to a narrow sound profile. Parameter modulation substantially improves generalization, especially with brightness and detune having the largest impact. The combined modulation achieves the highest F1-Score ($\SI{64.44}{\percent}$), tripling recall while maintaining precision. These results emphasize the importance of timbral variability over strict audio fidelity.

\begin{table}
  \caption{Impact of individual Karplus-Strong synthesis parameter modulations on transcription performance. Metrics (Precision, Recall, F1-Score) are reported on the full test set. Each row isolates a single modulated parameter, while the "Combined" row reflects the use of all modulations together, demonstrating their cumulative effect on model generalization.}
  \label{tab:synthesis_results}
  \centering
  \begin{tabular}{lccc}
    \toprule
    \textbf{Parameter Modulation} & \textbf{Precision}     & \textbf{Recall}        & \textbf{F1-Score}      \\
    \midrule
    None                          & $\SI{70.16}{\percent}$ & $\SI{18.02}{\percent}$ & $\SI{27.65}{\percent}$ \\
    \midrule
    Amplitude                     & $\SI{58.30}{\percent}$ & $\SI{24.63}{\percent}$ & $\SI{33.54}{\percent}$ \\
    Brightness                    & $\SI{71.08}{\percent}$ & $\SI{30.53}{\percent}$ & $\SI{41.25}{\percent}$ \\
    Level                         & $\SI{70.49}{\percent}$ & $\SI{21.51}{\percent}$ & $\SI{31.73}{\percent}$ \\
    Position                      & $\SI{74.28}{\percent}$ & $\SI{20.48}{\percent}$ & $\SI{30.88}{\percent}$ \\
    Detune                        & $\SI{79.65}{\percent}$ & $\SI{53.18}{\percent}$ & $\SI{61.67}{\percent}$ \\
    \midrule
    Combined                      & $\SI{70.62}{\percent}$ & $\SI{61.42}{\percent}$ & $\SI{64.44}{\percent}$ \\
    \bottomrule
  \end{tabular}
\end{table}

\subsection{Effect of audio augmentation}

Since the Karplus-Strong model simulates only string excitation, we evaluated the role of post-processing audio effects in simulating realistic recordings. Building upon the combined modulation setup, we applied each augmentation individually and in combination.

As shown in Table~\ref{tab:augmentation_results}, individual augmentations yield modest improvements. However, combining all effects (distortion, filtering, reverb, and noise) produces a noticeable performance boost, particularly for solos, where the F1-Score improves from $\SI{70.35}{\percent}$ to $\SI{77.02}{\percent}$ (not shown in Table \ref{tab:augmentation_results}). This highlights the value of environmental realism in audio generation.

\begin{table}
  \caption{Evaluation of individual and combined audio augmentation strategies applied to Karplus-Strong synthesized training data. Results are reported as Precision, Recall, and F1-Score on the full test set, illustrating the contribution of each effect (distortion, filtering, reverb, noise) to transcription performance and the cumulative benefit of combined augmentation.}
  \label{tab:augmentation_results}
  \centering
  \begin{tabular}{lccc}
    \toprule
    \textbf{Augmentation} & \textbf{Precision}     & \textbf{Recall}        & \textbf{F1-Score}      \\
    \midrule
    None                  & $\SI{67.29}{\percent}$ & $\SI{63.79}{\percent}$ & $\SI{64.41}{\percent}$ \\
    \midrule
    Distortion            & $\SI{70.09}{\percent}$ & $\SI{60.85}{\percent}$ & $\SI{63.94}{\percent}$ \\
    Lowpass               & $\SI{68.95}{\percent}$ & $\SI{60.84}{\percent}$ & $\SI{63.43}{\percent}$ \\
    Highpass              & $\SI{68.81}{\percent}$ & $\SI{62.84}{\percent}$ & $\SI{64.67}{\percent}$ \\
    Reverb                & $\SI{75.36}{\percent}$ & $\SI{57.81}{\percent}$ & $\SI{64.38}{\percent}$ \\
    Noise                 & $\SI{72.28}{\percent}$ & $\SI{61.23}{\percent}$ & $\SI{64.79}{\percent}$ \\
    \midrule
    Combined              & $\SI{80.62}{\percent}$ & $\SI{60.83}{\percent}$ & $\SI{68.22}{\percent}$ \\
    \bottomrule
  \end{tabular}
\end{table}

\subsection{Procedural data generation performance}

\begin{table}[htb]
  \caption{Evaluation results (in percent) for models trained on procedurally generated datasets composed using different tablature composition methods. Metrics include Precision (P), Recall (R), and F1-Score (F1), reported separately for full GuitarSet test data, accompaniments (Comp), and solos (Solo).}
  \label{tab:procedural_results}
  \centering
  \begin{tabular}{l|ccc|ccc|ccc}
    \toprule
    \textbf{Composer} & \multicolumn{3}{c|}{\textbf{Full}} & \multicolumn{3}{c|}{\textbf{Comp}} & \multicolumn{3}{c}{\textbf{Solo}}                                                                                 \\
    \cmidrule{2-10}
                      & \textbf{P}                         & \textbf{R}                         & \textbf{F1}                       & \textbf{P} & \textbf{R} & \textbf{F1} & \textbf{P} & \textbf{R} & \textbf{F1} \\
    \midrule
    Simple            & 70.11                              & 46.65                              & 52.64                             & 73.81      & 28.95      & 40.37       & 66.41      & 64.35      & 64.91       \\
    MMM               & 68.68                              & 55.30                              & 60.21                             & 68.39      & 46.35      & 54.28       & 68.97      & 64.24      & 66.14       \\
    Fingerpicking     & 74.69                              & 61.99                              & 66.23                             & 73.79      & 47.08      & 56.42       & 75.59      & 76.90      & 76.04       \\
    \bottomrule
  \end{tabular}
\end{table}

\begin{figure}[ht]
\tikzstyle{input} = [rectangle, rounded corners, text width=3em, minimum height=2em, text centered, draw]
\tikzstyle{output} = [rectangle, rounded corners, text width=3em, minimum height=2em, text centered, draw]
\tikzstyle{block} = [rectangle, draw, text centered, text width=6em, minimum height=1.6em]
\tikzstyle{decision} = [diamond, draw, text centered, text width=4em, minimum height=1.6em]
\tikzstyle{line} = [thick,->,>=stealth]

\begin{tikzpicture}
  \node [input] (start) {Start};
  \node [block, right=1cm of start, text width=4em] (move) {Move\\Time};
  \node [block, right=1cm of move, text width=8em] (insert) {Insert Notes for\\each String};
  \node [decision, right=1cm of insert] (length) {Long\\Enough?};
  \node [output, right=1cm of length] (end) {End};

  \draw [->] (start) -- (move);
  \draw [->] (move) -- (insert);
  \draw [->] (insert) -- (length);
  \draw [->] (length) -- (end);
  \draw [->] (length.south) |- ([yshift=-0.5cm]length.south) -| ([xshift=-0.5cm]move.west);
  \node at ($(length.east)+(0.25cm, 0.25cm)$) {Yes};
  \node at ($(length.south)+(-0.3cm, -0.25cm)$) {No};
\end{tikzpicture}
  \centering
  \caption{Flowchart of the greedy random tablature generation algorithm.}
  \label{fig:simple_composer_flow}
\end{figure}
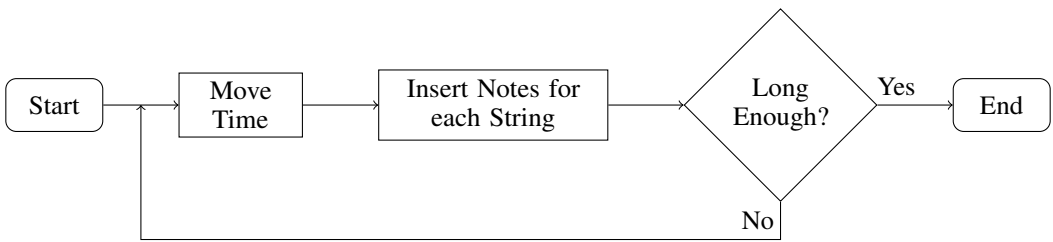

In this experiment, we evaluate the full procedural pipeline's capability to generate useful training data from scratch, without relying on existing tablatures. We compare three data generation strategies:

\begin{enumerate}
  \item \textbf{Simple Greedy Generator}: A naive algorithm that iteratively inserts random notes for each string with randomly sampled durations until a fixed target length is reached (see Figure~\ref{fig:simple_composer_flow}). This method does not incorporate any harmonic, rhythmic, or stylistic constraints.
  \item \textbf{MMM Transformer Model}~\citep{ens2020mmm}: A neural generative model trained on quantized note sequences from the GuitarSet training split. The model is used to generate MIDI representations, which are then converted to tablature and rendered into audio using the synthesis and augmentation pipeline.
  \item \textbf{Proposed Fingerpicking Composer}: A rule-based system that samples realistic fingerstyle patterns from a curated database and applies them to structured chord progressions. The compositions are transposed, humanized, rendered to audio using Karplus-Strong synthesis, and finally augmented to emulate realistic performance conditions.
\end{enumerate}

The comparative results are presented in Table~\ref{tab:procedural_results}. Among the three methods, the simple generator performs the worst, especially for accompaniment tracks, where its lack of structure and musicality results in poor recall and overall low transcription quality. The MMM transformer offers a significant improvement, particularly for comping, due to its data-driven understanding of harmonic and rhythmic structure. However, its results still lag behind those achieved with real or structured synthetic data.

Our proposed fingerpicking composer outperforms both baselines across all metrics and subsets. It achieves a notable $\SI{15}{\percent}$ relative increase in F1-Score on solo tracks compared to the MMM model and an even greater improvement over the simple generator. This demonstrates the importance of musical structure and stylistic relevance in synthetic training data for transcription tasks.

To further investigate the components contributing to the performance of the fingerpicking pipeline, we conduct an ablation study isolating the effects of audio augmentation and MIDI humanization.

\begin{table}
  \caption{Ablation study evaluating the impact of humanization and audio augmentation on transcription performance. Metrics (Precision, Recall, F1-Score) are reported for the full test split. The proposed method includes both techniques; ablated versions omit either augmentation or humanization to assess their individual contributions.}
  \label{tab:humanization_results}
  \centering
  \begin{tabular}{lccc}
    \toprule
    \textbf{Ablation} & \textbf{Precision}     & \textbf{Recall}        & \textbf{F1-Score}      \\
    \midrule
    Proposed          & $\SI{74.69}{\percent}$ & $\SI{61.99}{\percent}$ & $\SI{66.23}{\percent}$ \\
    No Augmentation   & $\SI{61.38}{\percent}$ & $\SI{56.95}{\percent}$ & $\SI{57.54}{\percent}$ \\
    No Humanization   & $\SI{79.18}{\percent}$ & $\SI{52.36}{\percent}$ & $\SI{61.90}{\percent}$ \\
    \bottomrule
  \end{tabular}
\end{table}

As shown in Table~\ref{tab:humanization_results}, removing audio augmentation results in a $\SI{15}{\percent}$ drop in F1-Score, underscoring the importance of simulating recording imperfections and environmental acoustics. Similarly, removing the humanization step, responsible for introducing small timing and pitch variations, yields a $\SI{7}{\percent}$ performance reduction. These variations likely improve model robustness to expressive nuance in real recordings.

Taken together, these findings validate our full procedural pipeline as an effective approach for generating realistic and diverse training data. While the fingerpicking composer captures the musical essence of fingerstyle guitar, it is the combination with expressive synthesis and augmentation that enables generalization to real-world recordings.

Nevertheless, the results also highlight that the main performance bottleneck lies not in the composition, but in the fidelity of the audio rendering. Future improvements in physical modeling or differentiable synthesis could help bridge the remaining performance gap with real recordings.

To illustrate the diversity and realism achieved through our pipeline, we provide a curated set of audio examples covering the full spectrum of generated data\footnote{\url{https://github.com/klangio/procedural-data-training}}. This includes excerpts from the three compositional strategies (simple, MMM, and fingerpicking-based), as well as side-by-side comparisons of different synthesis settings and augmentation effects.

\subsection{Finetuning with real audio}

\begin{table}[htb]
  \caption{Transcription performance after finetuning on real recordings, comparing models trained from scratch on real data, trained on procedural data only, and pretrained on procedural data followed by finetuning. Results are reported in percent as Precision (P), Recall (R), and F1-Score (F1) for the full test set, accompaniment (Comp), and solo (Solo) subsets.}
  \label{tab:finetuning_results}
  \centering
  \begin{tabular}{l|ccc|ccc|ccc}
    \toprule
    \textbf{Training Data} & \multicolumn{3}{c|}{\textbf{Full}} & \multicolumn{3}{c|}{\textbf{Comp}} & \multicolumn{3}{c}{\textbf{Solo}}                                                                                 \\
    \cmidrule{2-10}
                           & \textbf{P}                         & \textbf{R}                         & \textbf{F1}                       & \textbf{P} & \textbf{R} & \textbf{F1} & \textbf{P} & \textbf{R} & \textbf{F1} \\
    \midrule
    Real Recordings        & 92.05                              & 73.84                              & 81.42                             & 90.75      & 63.02      & 74.15       & 93.35      & 84.66      & 88.70       \\
    Procedural Data        & 74.69                              & 61.99                              & 66.23                             & 73.79      & 47.08      & 56.42       & 75.59      & 76.90      & 76.04       \\
    Finetuning             & 92.14                              & 76.95                              & 83.49                             & 90.44      & 68.02      & 77.41       & 93.85      & 85.87      & 89.56       \\
    \bottomrule
  \end{tabular}
\end{table}

To investigate the utility of procedural data for pretraining, we first trained models for $10,000$ steps on procedurally generated audio, then finetuned on real recordings. As shown in Table~\ref{tab:finetuning_results}, pretraining yields a modest $\SI{2}{\percent}$ F1-Score gain over training solely on real data.

\begin{figure}[ht]
  \begin{tikzpicture}
  \begin{axis}[
      ybar,
      symbolic x coords={
          1/5 Splits,
          2/5 Splits,
          3/5 Splits,
          4/5 Splits,
          5/5 Splits
        },
      xtick=data,
      enlarge x limits=0.1,
      ylabel=F1-Score (\%),
      ymin=0,
      ymax=100,
      bar width=0.5cm,
      width=12cm,
      height=6cm,
      legend style={at={(0.5,-0.15)},
          anchor=north,legend columns=-1},
      legend image code/.code={
          \draw [#1] (0cm,-0.1cm) rectangle (0.2cm,0.25cm); },
    ]
    \addplot[fill=highlight1]
    coordinates {
        (1/5 Splits,63.32)
        (2/5 Splits,71.03)
        (3/5 Splits,77.20)
        (4/5 Splits,79.63)
        (5/5 Splits,81.42)
      };
    \addplot[fill=highlight2]
    coordinates {
        (1/5 Splits,77.45)
        (2/5 Splits,79.22)
        (3/5 Splits,81.32)
        (4/5 Splits,82.32)
        (5/5 Splits,83.49)
      };
    \legend{w/o Pre-Training, w/ Pre-Training}
  \end{axis}
\end{tikzpicture}
  \centering
  \caption{Demonstration of the effectiveness of pre-training when reducing the amount of real data. The training dataset is divided by guitarist into five splits with 60 recordings each. The evaluation is performed on the full test split.}
  \label{fig:reduced_data_chart}
\end{figure}
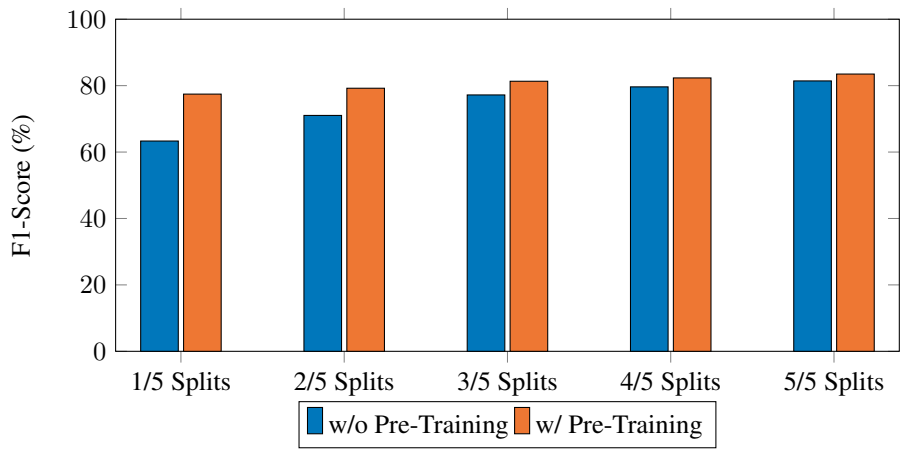

The benefits become more pronounced with reduced finetuning data. Figure~\ref{fig:reduced_data_chart} illustrates that pretraining enables comparable performance with only a fifth of the real recordings. For example, training from scratch with $60$ recordings results in an F1-Score of $\SI{63.32}{\percent}$, whereas pretraining lifts that to $\SI{77.45}{\percent}$.

This demonstrates the value of procedural pretraining, especially in low-data scenarios or for underrepresented instruments that could benefit from similar synthesis pipelines.
\section{Conclusion}
\label{sec:conclusion}
In this work, we investigated the use of procedurally generated training data for the task of automatic transcription of fingerpicked acoustic guitar performances. We introduced a novel data generation pipeline comprising the composition of fingerpicking tablatures, conversion to performance-level MIDI, audio synthesis using an extended Karplus-Strong algorithm, and audio augmentation. This fully synthetic pipeline enables the creation of annotated training data without reliance on copyrighted recordings.
Our experiments demonstrated that models trained solely on procedurally generated audio can achieve competitive transcription accuracy. Moreover, pretraining on synthetic data followed by finetuning on real recordings yielded improved performance compared to training exclusively on real data and requires significantly fewer real recordings to achieve comparable results. These findings suggest that procedural data generation can be a powerful tool for overcoming data scarcity in music transcription tasks.

Future work could explore applying this procedural training approach to more advanced model architectures such as transformer-based models \citep{gardner2022mt3} or regression-based CRNNs \citep{riley2024high}. Additionally, integrating differentiable digital signal processing (DDSP) synthesis techniques \citep{jonason2023ddsp} could enable richer supervision, such as direct prediction of string and fret positions. Beyond guitar transcription, the procedural generation framework could be adapted to other instruments, offering a scalable solution for tasks with limited annotated real-world data.





\bibliographystyle{apalike}
\bibliography{aimc2025_procedural_data}

\end{document}